# Bessel beam fabrication of graphitic micro electrodes in diamond using laser bursts

Akhil Kuriakose[a,b], Francesco P. Mezzapesa[c], Caterina Gaudiuso[c], Andrea Chiappini[d], Federico Picollo[e], Antonio Ancona[c] and Ottavia Jedrkiewicz[a]

a. Istituto di Fotonica e Nanotecnologie (IFN)-CNR, Udr di Como, Via Valleggio 11, 22100 Como, Italy; akuriakose@studenti.uninsubria.it, ottavia.jedrkiewicz@ifn.cnr.it

b. Dipartimento di Scienza e Alta Tecnologia, Università dell'Insubria, Via Valleggio 11, 22100 Como, Italy;

c. Istituto di Fotonica e Nanotecnologie (IFN)-CNR, Dipartimento Interateneo di Fisica, Università degli Studi di Bari, Via Amendola 173, 70125 Bari, Italy;

francesco.mezzapesa@ifn.cnr.it; caterina.gaudiuso@ifn.cnr.it; antonio.ancona@uniba.it

d. Istituto di Fotonica e Nanotecnologie (IFN)-CNR, CSMFO Lab and FBK Photonics Unit, Via alla Cascata 56/c, Povo, 38123 Trento, Italy; andrea.chiappini@ifn.cnr.it

e. Department of Physics and "NIS" Inter-departmental Centre, University of Torino National Institute of Nuclear Physics, sect. Torino, Via Pietro Giuria 1, 10125, Torino, Italy;

Correspondence: ottavia.jedrkiewicz@ifn.cnr.it

# Abstract

We present the fabrication of conductive graphitic microelectrodes in diamond by using pulsed Bessel beams in the burst mode laser writing regime. The graphitic wires are created in the bulk of a 500 μm thick monocrystalline HPHT diamond (with (100) orientation) perpendicular to the sample surface, without beam scanning or sample translation. In particular, the role of different burst features in the resistivity of such electrodes is investigated for two very different sub-pulse durations namely 200 fs and 10 ps, together with the role of thermal annealing. Micro-Raman spectroscopy is implemented to investigate the laser-induced crystalline modification, and the results obtained by using two different laser repetition rates, namely 20 Hz and 200 kHz, are compared. A comparison of the micro-Raman spectra and of the resistivity of the electrodes fabricated respectively with 10 ps single pulses and with bursts (of sub-pulses) of similar total duration has also been made, and we show that the burst mode writing regime allows to fabricate more conductive micro electrodes, thanks to the heat accumulation process leading to stronger graphitization. Moreover, the microfabrication of diamond by means of the longest available bursts (~ 46.7 ps duration) featured by 32 sub-pulses of 200 fs duration, with intra-burst time delay of 1.5 ps (sub-THz bursts), leads to graphitic wires with the lowest resistivity values obtained in this work, especially at low repetition rate such as 20 Hz. Indeed, micro electrodes with resistivity on the order of 0.01 Ω cm can be fabricated by Bessel beams in the burst mode regime even when the bursts are constituted by femtosecond laser sub-pulses, in contrast with the results of the standard writing regime with single fs pulses typically leading to less conductive micro electrodes.

# 1. Introduction

Fabrication of microstructures in diamond for various novel applications such as radiation detectors [1], microfluidic chips [2] and photonic circuits [3] is attracting a lot of attention thanks to the unique properties of this crystal such as high scratch resistance, remarkable hardness with a Mohs index of 10, increased chemical resistivity, high radiation stability and high thermal conductivity [4, 5]. Thanks to its good biocompatibility, diamond is used as a strong candidate for biosensing as well [6]. One of the high impact works that is being investigated widely is the incorporation of Nitrogen Vacancy (NV) centres in the diamond bulk for quantum communication and nanoscale magnetometry [2]. A common feature in all these applications is the presence of graphitic electrodes which act as conductive pathways for different kinds of sensors. The main techniques to fabricate such conductive channels are annealing [7] where the electrodes are fabricated mainly on the surface of the diamond substrate, focused ion beam [8, 9] which specialises in fabrication within the bulk but close to the surface depending on the energy of the ion beam, and laser writing [10-14] which gives freedom to write in any geometry at any depth with desired length. Therefore, laser writing can also be used for designing and fabricating 3D structures as well within the diamond bulk.

The precise modification of different transparent materials within a confined volume is possible thanks to the use of ultrafast pulsed lasers. The non-linear absorption processes of intense ultrashort pulses in the material reduce the collateral damages due to thermal effects thus helping in pin-pointing the modification without destroying the whole sample [15]. It is thus possible to write different microstructures in any configurations with any dimensions both on the surface and in the bulk making use of such lasers. As for diamond which is insulating in nature, a conventional focused Gaussian beam with sufficient fluence can be used for the optical breakdown of the material thus transforming locally the diamond into graphite. The whole process includes the primary optical breakdown of the material which results in the formation of discrete globules of graphite in the nearest vicinity of the laser focus and the propagation of a graphitization wave from those globules to the laser focus. By translating the beam (or the sample stage) throughout the sample, one can fabricate a wire-like graphitized region with a length of our choice where the diameter of the wire is close to the beam spot size [10,16]. While the fabrication using Gaussian beams requires the translation of stage or the beam scanning for the creation of long straight electrodes, with the use of Bessel beams such movements are not needed. Finite energy Bessel beams (i.e., Bessel–Gauss beams) are characterised by a cylindrical symmetry and a transverse profile featured by an intense central spot surrounded by weaker concentric rings [17]. Their non-diffracting zone, invulnerable to the diffraction spreading in contrast to Gaussian beams [18], has been utilised for in-bulk high aspect-ratio modifications of transparent materials without the need for sample translation [19], even in single shot [20]. It is worth noting that Bessel beams find their applications not just in internal microstructuring of transparent materials but also for other technological applications such as drilling, cutting or welding thanks to their self-reconstruction property and the elongated focal zone [21-24].

The Bessel fabrication of graphitic electrodes perpendicular to the diamond sample surface has already been presented in recent works by our group. While a first study has led on one hand to a top-notch resistivity value of 0.04 Ω cm in (100) oriented samples [25], the exploration of the role of the crystallographic orientation of the diamond in the electrodes conductivity has shown that those fabricated in a (110) cut sample are more conductive compared to those fabricated in the former one, thanks to the additional heating mechanism associated with electrode growth in the (110) orientation case. This resulted in better transformation of $sp^3$ diamond into $sp^2$ structures, bringing down the electrodes resistivity value to 0.013 Ω cm [26].

Even though the use of ultrafast lasers helps in reducing the collateral damages in the material due to thermal effects, heat dissipation occurring in some laser machining regimes can lead to a reduction in the high quality and efficiency of the fabrication. In particular, the laser energy deposited onto the sample in the form of heat while using the standard single-pulse mode laser operation (i.e train of single pulses) can still lead to a few undesired damages such as cracks or melting, which can, in turn, worsen the quality of the microstructures, as also observed when creating micro holes [27]. More importantly, it also causes the diffusion of heat from the laser focal zone thus reducing the effective modification in the intended region. This energy dissipation can also lead to a rapid heating and cooling cycle, again producing potentially detrimental effects such as micro cracks [28]. In order to counter such outcomes, one of the most prominent approaches is to deliver laser pulses in bursts (for the pioneering work on bursts see [29]) where residual thermal energy does not have sufficient time to diffuse out of the laser interaction zone before the arrival of the next laser pulse [30]. The burst mode processing consists of using sequences (or bursts) of several ultrashort laser pulses with inter-pulse time delay shorter than 1 μs. This helps in stopping the heat from seeping out by setting up a thermal modification zone within the focus of the beam itself thus helping in an efficient modification. Heat accumulation effects over such burst trains have already been found beneficial in various applications such as fabrication of deep high-aspect ratio holes in metals [31] and dielectrics [32] and of hierarchical structures conferring superhydrophobic wetting response to copper [33].

In the processing of materials by laser bursts, a single pulse (i.e. each pulse exiting from the laser source at the nominal repetition rate) is replaced by a pulse package including several sub-pulses with equally fractioned energy at an intra-burst frequency much higher than the laser repetition rate that is typically in the KHz to the MHz regime. The time delay between sub-pulses within the bursts can be as short as a few picoseconds, meaning that the intra-burst frequency is in the sub-TH regime. The advantage of using such laser bursts is that the target material can undergo efficient modification before the residual heat deposited by previous pulses diffuses away from the processing region thus increasing the effectual transformation which can be either a structural or phase modification. The burst laser systems are generally used for material processing especially in metals [34]. For instance, Karmer et al. [35] showed that by using burst modes one can benefit from increased material removal rate if the peak fluence of each pulse within the burst is nearer the optimum value needed for the maximum specific removal rate. Gaudiuso et al. [36] have reported that the main factors affecting the ablation process are the average power, the scan speed and the number of pulses within the burst. Even though the advantage of using bursts of pulses was not that evident in terms of ablation removal rate, the burst mode allowed a fine tuning of the surface finishing in micro-milling of steel. In other words, when shallow milling depths are needed, using bursts with high number of pulses is the most reliable solution to ensure high quality of the surface finishing, i.e. reduced surface roughness.

Here, in this work, we try to take advantage of the heat stack up nature of bursts to initiate better transformation of diamond into amorphous carbon thus improving the conductivity of electrodes fabricated by Bessel beams in a (100)-cut sample. The aim of the work is to establish the importance of the additional heating mechanism occurring in a burst writing regime for the laser fabrication of highly conductive electrodes. Different burst configurations will be explored to understand the correlation between conductivity and burst parameters. A direct comparison between single pulse mode and burst mode writing regimes will be established by reporting the resistivity values of electrodes fabricated with same laser parameters. In addition, this work also highlights the role of the repetition rate in the resistivity of the laser-written graphitic wires. Using the best combination of burst configuration and laser writing parameters, we aim to improve the resistivity values previously obtained for electrodes fabricated in diamond with (110) orientation [26].

# 2. Materials and Methods

*2.1 Samples details*

To carry out an extensive study of the micromachining of diamond using various laser bursts configurations, two samples of monocrystalline type IIa HPHT diamond (obtained with High Pressure High Temperature synthesis thecnique) with (100) orientation are considered. While one sample is completely dedicated to the electrodes fabricated with bursts, the second one is used for single pulse fabrication so that a thorough comparison can be done. Both samples were provided by MB Optics. The sample dimensions are 5 mm × 5 mm × 0.5 mm. All the surfaces including the side-edges are polished so that the lateral analysis of the electrodes through optical microscopy is made possible.

*2.2 Experimental set-up*

The laser source used in this experiment is the Ytterbium laser Pharos-SP from Light Conversion Ltd. delivering a linearly polarized beam at a wavelength of 1030 nm and a tunable pulse duration from 200 fs to 10 ps. The maximum available average power and pulse energy are given by 6 W and 1.5 mJ respectively, and the repetition rate can be tuned between 1 kHz and 1 MHz. In addition, a pulse picker allows selecting the number of pulses exiting from the laser source in order to reduce the repetition rate down to 20 Hz, for a comparison with previous results obtained at this pulse frequency [25,26].

The Bessel beam is obtained by sending a spatially filtered Gaussian beam with a ~ 4 mm total diameter through a fused silica axicon with 178º apex angle. A telescopic system formed by a 229 mm focal length lens ($L_1$) and a 0.45 N.A. 20× microscope objective ($f_{obj}$ = 10 mm), leads to a demagnified Bessel beam featured by 12º cone angle, a total central core size of about 2 µm and a total Bessel zone of about 600 µm at the sample position, characterized prior to micromachining with a one lens imaging system and a CMOS camera (Gentec BEAMAGE-4M); for details on the typical intensity distribution see for instance [37]. The energy is set and varied using an energy controller consisting of a half wave plate and a polarizer, and was chosen in a such a way to work, for each writing mode, in the optimal range leading to micro electrodes with lowest resistivity. The Bessel beam is injected orthogonally to the sample as illustrated in the experimental set up schemes reported in [25,26], and thanks to its elongated focal zone covering the entire 500 µm thickness of the diamond, the fabrications are carried out without any sample translation. A real-time observation of the sample surfaces during the micromachining process is realized thanks to a backlighting Light Emitting Diode (LED) placed below the transparent sample and an imaging system using a lens and another CCD.

This allows a careful adjustment of the relative positioning of the Bessel zone with respect to the sample thickness. By verifying that the top and bottom surface damage traces obtained in single shot are of similar shape and size, we can ensure for the case of in-bulk microfabrication that the central core peak intensity along the Bessel zone is symmetrically distributed across the sample, i.e., maximum intensity in the middle of the sample, and smaller intensity at the top and bottom. This also prevents strong input interaction at the air/diamond interface where beam reshaping occurs because of the sudden change in the refractive index.

In order to generate the internal graphitic wires covering the whole sample thickness, the micro fabrications are carried out in the multiple shot regime, as the length of the obtained micro electrode depends on the number of laser pulses irradiating the diamond. In the single pulse writing regime with 200 fs pulses (where the growth turned out to be slower) it was found that a minimum of 7000 Bessel pulses is enough to create an electrode of 500 μm length, provided that the Bessel thickness zone covers the entire diamond [25]. All the experiments performed in this work have been carried out with 9000 laser pulses or bursts. Finally note that the threshold fluence of our Bessel core for having surface damages in single shot is in the range of 1-2 μJ/cm$^2$ (the lower value in the ps laser pulsed regime) while for the internal graphitic wire growth in multiple shot a double fluence is typically needed.

For the implementation of the burst mode system, following pioneering works that demonstrated the possibility to temporally shape ultrashort pulses with nonlinear crystals [38; 39], the beam line has been modified by adding five birefringent calcite crystals at the laser output [33, 36]. In this experimental configuration, before the Bessel beam shaping, each incoming laser pulse is split into 2 sub-pulses by each crystal, for a total number of up to n=2$^5$ sub-pulses within the bursts, depending on the mutual orientation of the pulses polarization and the optical axes of the calcite crystals. The intra-burst frequency (in sub-THz regime in our case) also depends both on the crystal orientation and thickness. The pulses within the burst (i.e., the sub-pulses) have alternate crossed polarizations, half of them along the optical axes of the crystals. Bursts with different number of sub-pulses (n = 2, 8 and 32) ar used in our fabrication experiments and their time delay Δt is varied between 1.5 ps and 46.5 ps, while working with pulse durations of 200 fs. When employing 10 ps pulse widths, we use bursts with n=2 sub-pulses having a time delay Δt=36 ps or Δt =46.5 ps. While the details of all burst configurations used in this work are presented in Table 1, a schematic of one of those (for instance mode 3) is shown for illustration in Figure 1. It is worth noting that the temporal duration of all the available bursts lies in the picosecond range, while the laser sub-pulses within the burst packages can have a duration of 200 fs or 10 ps.

|  | **200 fs (sub-pulse duration)** | | | | **10 ps (sub-pulse duration)** | |
|---|---|---|---|---|---|---|
|  | **Mode 1** | **Mode 2** | **Mode 3** | **Mode 4** | **Mode 5** | **Mode 6** |
| **n** | 32 | 2 | 8 | 2 | 2 | 2 |
| **Δt** | 1.5 ps | 46.5 ps | 1.5 ps | 10.5 ps | 46.5 ps | 36 ps |
| **τ$_B$** | 46.7 ps | 46.7 ps | 10.7 ps | 10.7 ps | 56.5 ps | 46 ps |

Table 1: Parameters of the laser burst configurations used (n is the number of sub-pulses within the burst, Δt is the time delay between these, and τ$_B$ is the total burst duration). The laser pulse duration (duration of each sub-pulse) was set to be either 200 fs or 10 ps.

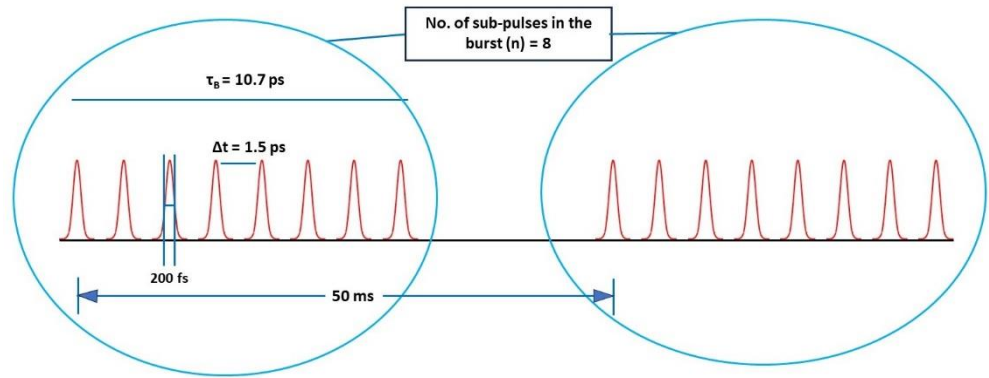

Figure 1: Schematic of the burst mode 3 described in Table 1, drawn for a 20 Hz repetition rate. Note that in case of 200 kHz repetition rate the time distance between bursts is 5 μs.

## 2.3 Characterization of the graphitic micro electrodes

Optical microscopy has been used to observe the graphitic micro wires fabricated close to the lateral edges of the (100) diamond sample, confirming on one hand the same type of pulse (in this case burst) duration and energy dependences of the morphology already observed in [25]; and on the other hand, highlighting that the morphology of the laser written electrode, with central core size of ~ 2-3 μm, does not depend on the different infrared wavelength used here (i.e. 1030 nm), nor on the operating mode of the laser, i.e. delivering either separated single pulses or bursts of pulses.

Micro-Raman spectroscopy has been used to characterize the crystalline modification of the material after laser irradiation, and because the measurement on the central portions of the graphitic wires embedded in the diamond bulk leads to a too strong diamond peak contribution and a heavily suppressed G-band, it has been performed at one of the graphitic wire extremities in correspondence with the sample surface. This has been done by means of a Labram Aramis John Yvon Horiba system with a diode-pumped solid state (DPSS) laser source at 532 nm, also equipped with a confocal microscope, a 1800 line mm$^{-1}$ grating and an air-cooled CCD camera. The nominal power on the sample was about 4 mW. The Raman signal collected with a 100× objective in a back-scattering configuration is featured by a spatial resolution of about 1 μm. The laser beam used for the measurement has been focused at the centre of the transverse section of the wires extremity (a few micrometer below the sample surface in order to maximize the signal), where we expect a higher concentration of graphite, as previously discussed [25]. Note that although these measurements allow to evaluate the crystalline structure and modification of the sample only close to the surfaces, the comparative results can still give an indication of the different types of transformation occurring in different fabrication regimes, and can be used as a preliminary tool to understand the different graphitic contents in the electrodes obtained with different laser parameters.

Finally, the electrical characterization of the fabricated micro electrodes has been carried out, as described in [25,26], by using conductive layers on the top and bottom of the diamond samples. In particular, silver paste was applied on the top surface covering the whole area as a small ablation crater of few micrometres depth maximum (when using high intensities) is formed in correspondence with the extremity of the graphitic wire, while on the other side where the crater is very shallow a silver layer was deposited thanks to a mask which was created by milling an aluminium film with 15 μm of thickness, in order to isolate one electrode from the other and avoid short circuit during the current intensity-voltage (IV) electrical measurements. The IV tests have been repeated after thermal annealing of the micromachined sample (carried out at $10^{-2}$ mbar pressure and 950 degree Celsius for 1 hour) only for the electrodes featured by a potential barrier and for the electrodes fabricated in the burst configuration featured by the lowest resistivity values. Unless specified, all the resistivity values and IV curves reported in this paper are measured before annealing. The resistivity ρ is obtained using the equation $\rho = RA/L$, where the resistance $R=V/I$ is calculated in a quite standard way by using the linear portion of the IV curves and considering the maximum intensity value (and the corresponding voltage value) before saturation; A is the cross-sectional area of the electrode core (always featured by a diameter of about 2.5 μm, apart from the electrode relative to the measurement shown in Fig. 6, where it was 3 μm), and L is the total length of the electrode (500 μm). Finally, note that the standard deviation error associated with the measurement is on the order of 5%.

# 3. Graphitic electrode fabrication with femtosecond sub-pulses

In this section we present the results related to the experiments for the Bessel beam fabrication of graphitic wires conducted with the same methodology presented in [25], and performed in multiple shot with laser bursts containing 200 fs sub-pulses, using both 20 Hz and 200 kHz repetition rates. As described in [25] the graphitic-like phase formation always starts at the bottom of the sample from a first graphitic damage seed formed at the air/sample interface and continues to grow in the direction of the Bessel beam until it reaches the top of the diamond sample.

## *3.1 Effect of different burst configurations*

The primary goal of our work is to understand the effect of different burst configurations on the resistivity of the graphitic micro wire obtained by Bessel beam micromachining. A first experiment has been performed by fabricating electrodes with three different bursts configurations, namely with bursts of 32 sub-pulses and total burst duration $\tau_B$= 46.7 ps (mode 1), with bursts of 8 sub-pulses and $\tau_B$ = 10.7 ps (mode 3) and finally with bursts of 2 sub-pulses and $\tau_B$ = 10.7 ps (mode 4), with 6 µJ per burst, and at two repetition rates namely, 20 Hz and 200 kHz. In figure 2 optical microscope images of the central portions of the graphitic wires fabricated in the different burst configurations are reported.

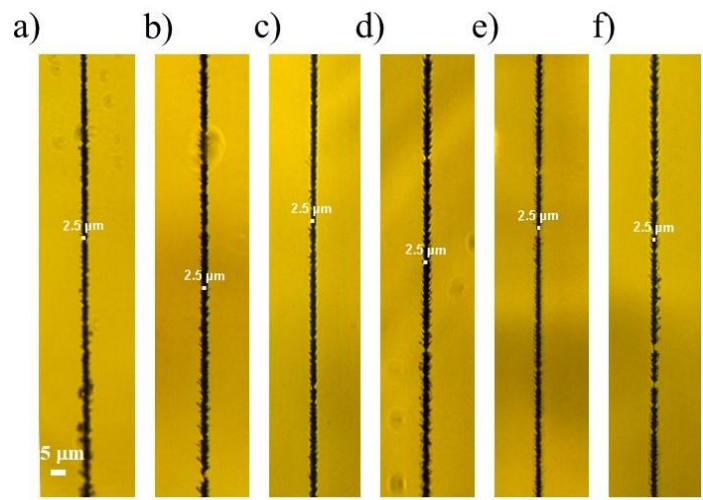

Figure 2: Zoomed optical microscope images of the central portion of the graphitic wires fabricated in different burst configurations namely a) mode 4 (200 kHz), b) mode 4 (20 Hz), c) mode 3 (200 kHz), d) mode 3 (20 Hz), e) mode 1 (200 kHz), f) mode 1 (20 Hz). The scale bar is the same for all images.

The results presented in figure 3 show that the writing configuration using bursts with larger number of sub-pulses, shorter intra-burst delay (sub-THz regime), and a longer total burst duration, leads to microelectrodes featured by lower resistivity compared to the other two configurations, for both 20 Hz and 200 kHz repetition rates. In accordance with these first results, in fact, we expected to have, for a given average power, better conductivity when using laser bursts featured by a large number of sub-pulses, as the heating process in the material (and thus the graphitization) during the radiation-matter interaction should have been more efficient compared to that occurring with a train of single pulses of the same duration (or even with bursts featured by few sub pulses).

A similar work has been reported in [40] where the interaction of a single 52 µJ laser pulse with soda lime glass led to a maximum temperature elevation of the order of 150 K which was not enough to reach the melting temperature and to induce permanent modifications. Meanwhile, a cumulative action of 4×13 µJ pulses in burst mode led to a peak temperature of 750 K in the central part of the beam thus modifying the material more efficiently. In the burst micromachining regime used here, each pulse in the burst interacts with the already heated diamond area (by a previous one), and as shown by our results, this additional heating mechanism due to the arrival of subsequent high repetition rate pulses within the bursts can have a considerable positive influence on the graphitization process, as also reported by Neuenschwander et al. [41] studying the burst mode modification of Diamond-like Nano composites; in particular as the number of sub-pulses increases, the interaction zone reaches a temperature way higher than the one that can be reached by single pulse machining. The same mechanism seems to apply to our work, leading to graphitic microelectrodes featured by very low resistivity values compared to those reported in [25] and close to the lowest resistivity value reported in [26] (in that case for electrodes fabricated in a (110) cut diamond). Moreover, note that the longer burst duration of the configuration 1 (i.e. mode 1) may also favour the graphitization process and thus the low resistivity of the resulting electrode, similarly to what reported in [25] where graphitic wires obtained using Bessel beam in the picosecond pulsed regime resulted in lower resistivity compared to those fabricated with femtosecond pulses.

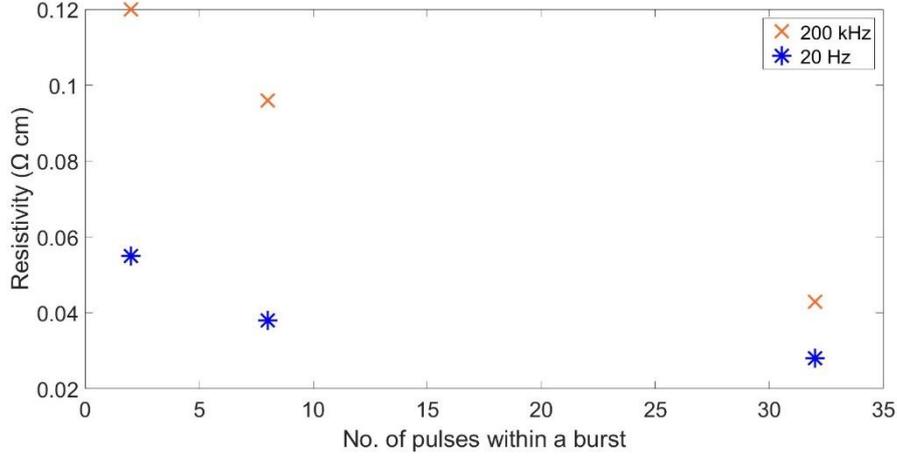

Figure 3: Evolution of the resistivity of microelectrodes fabricated in diamond by Bessel beam machining with three different burst configurations (mode 1 with 32 sub-pulses and $\tau_B$ = 46.7 ps, mode 3 with 8 sub-pulses and $\tau_B$ = 10.7 ps and mode 4 with 2 sub-pulses and $\tau_B$ = 10.7 ps, respectively), with 6 µJ energy per burst, 200 fs sub-pulse duration, and for two laser repetition rates namely 20 Hz and 200 kHz.

## 3.2 Effect of pulse repetition rate in burst mode writing

The results of Fig. 3 also highlight that irrespective of the burst configuration used, the micro electrodes fabricated with lower repetition rate are featured by lower resistivity than those obtained with higher laser repetition rate. A similar result was reported by Bharadwaj et al. [2]. From the Raman analysis performed on graphitic lines written with Gaussian beams repetition rates of 25 kHz and 5 kHz, at constant pulse energy (800 nJ), they observed that the G peak becomes sharper as the repetition rate is reduced, indicating an increased graphitization.

In order to confirm the same, micro-Raman measurements were conducted on the micro wires fabricated with laser bursts consisting of 2 sub-pulses (200 fs duration) and $\tau_B$ = 10.7 ps (mode 4) with 6 µJ energy per burst, and using two different repetition rates namely 20 Hz and 200 kHz.

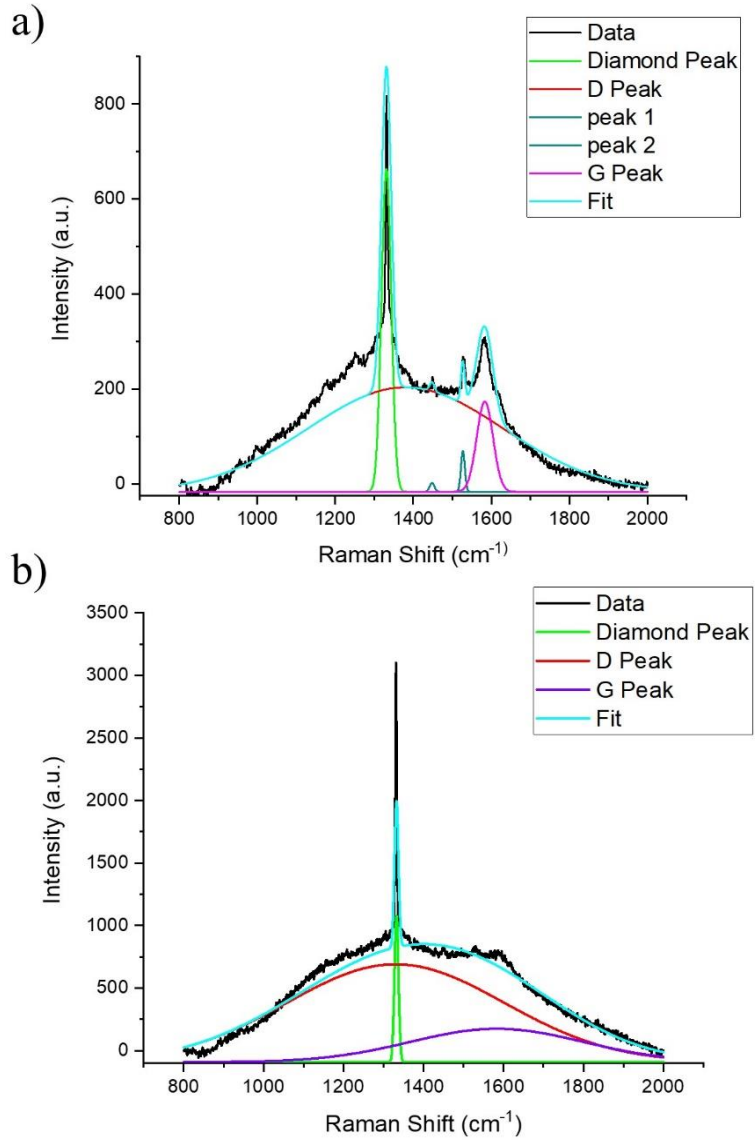

Figure 4: Micro-Raman spectra of two graphitic wires fabricated at two different repetition rates namely a) 20 Hz and b) 200 kHz using 6 µJ burst energy and 200 fs sub-pulse duration with the laser burst mode 4 (n=2 and $\tau_B$ = 10.7 ps).

Figure 4 presents the measured spectra obtained (black curve) and the "deconvolved" peaks considering the contribution of diamond and of amorphous carbon phases through the fitting procedure as described by A. Dychalska et. al. [42] that employed the use of a mix of Lorentzian and Gaussian functions, which are believed to represent the shape of a single band. In both cases, the presence of the D-band (around 1351 cm$^{-1}$) and G-band (around 1585 cm$^{-1}$) in addition to the diamond peak (1332.2 cm$^{-1}$) suggests that the Bessel beam micromachining at both repetition rates leads to a material conversion from sp$^3$ to sp$^2$ type bonding, thus transforming the diamond into a mixture of sp$^2$ carbon. At the same time, the rate of conversion which can be inferred from the intensity ratio of the G band to the diamond peak denotes that a better conversion of diamond into graphite occurs with low laser repetition rate. Indeed, not only the pure diamond peak intensity is lower for the spectrum of the micro wire fabricated at 20 Hz but, most importantly, the I(G)/I(Diamond) ratio in this case is around twice as that of the micro wire fabricated at 200 kHz. In addition, the G band looks sharper in case of 20 Hz fabrication where the width of the same is around one fifth of that of graphitic wire obtained with 200 kHz thus supporting our claim. Note that all other parameters were kept constant for this comparison apart from the repetition rate, and the highest amount of graphite in the micro electrode fabricated at 20 Hz with respect to that fabricated at 200 kHz is confirmed by the electrical measurements. Indeed, the evaluation of the electrodes resistivity from the IV measurements performed (curves not shown here) has led to a value of 0.055 Ω cm in the first case (low repetition rate fabrication) to be compared to the value of 0.1 Ω cm in the second (high repetition rate fabrication).

It is worth noticing that the spectrum of the electrode fabricated at 20 Hz (Fig. 4a) presents two evident additional bands around 1447 cm$^{-1}$ and 1528 cm$^{-1}$ respectively (highlighted by the dark green small peaks in the graph), compared to the spectrum of an electrode fabricated at 200 kHz with same parameters and same burst configuration, where the same are only weakly noticeable (Fig. 4b). While 1528 cm$^{-1}$ corresponds to sp$^1$ carbon [43], the role of 1447 cm$^{-1}$ has been debated in different works. In [44] the presence of 1447 cm$^{-1}$ as a band representing nanocrystalline diamond with sizes corresponding to 1–2 nm clusters has been highlighted, while Birrell et al. [45] ascribed the appearance of such band to the presence of hydrogen at the grain boundaries (carbon–hydrogen bonds) and so, not exactly sp$^3$ carbon. Finally, another work [46] contradicts the suggestion of [44] by reporting that this band should not be assigned to nanocrystalline diamond or other sp$^3$-bonded phases. The alternate suggestion is that these peaks can be assigned to transpolyacetylene segments (the first highly conductive organic polymer) at grain boundaries and surfaces and thus, they should be referred to sp$^2$ bonded carbon. Note that it has also been reported that the presence of a sp$^1$ (or simply sp) peak at 1528 cm$^{-1}$ peak in the Raman spectrum, indicates the semi-conductive to conductive nature of the modifications [47]. From the above considerations we believe that it may be assumed that the presence of these additional bands should not pose a deteriorating effect to the conductivity of our electrode. On the contrary, they could be possibly an indication of a better conductivity.

Finally, note that in our experiments the better transformation of diamond into graphite when using low repetition rate has been evaluated not just in the case of electrodes fabricated with laser bursts but also with single pulses. For instance, in the latter laser writing regime, an electrode obtained at 20 Hz, 200 fs and 6 μJ pulse energy yielded a resistivity of ≈ 0.5 Ω cm while its counterpart at 200 kHz (all other parameters being the same) showed a resistivity of ≈ 3.5 Ω cm (IV curves not shown here).

## 3.3 Effect of energy in burst mode writing

The effect of the energy in the micromachining process has been investigated by using the laser bursts configuration at 20 Hz that have led so far to the most conductive graphitic wires (i.e. mode 1 with n=32, $\tau_B$ = 46.7 ps, and a time delay of 1.5 ps between the sub-pulses). To this end, three different electrodes have been fabricated with decreasing burst energy using that particular configuration to understand the evolution of the resistivity of the micro electrodes. The results are reported in table 2, showing the decrease of the resistivity with increasing energy. Values after thermal annealing are also indicated in the table. In Figure 5 we present the IV graph from the electrical characterization after annealing of the electrode featured by the lowest resistivity obtained in this work, i.e. with a top-notch value of 0.01 Ω cm (fabricated with laser bursts in mode 1 and 10 µJ energy).

| Burst Energy (µJ) | Resistivity before annealing (Ω cm) | Resistivity after annealing (Ω cm) |
| --- | --- | --- |
| 10 | 0.012 | 0.01 |
| 8 | 0.02 | 0.019 |
| 6 | 0.028 | 0.024 |

Table 2: Evolution of the evaluated resistivity, with respect to the burst energy, of electrodes fabricated at 20 Hz in the laser burst mode 1 (with n = 32, Δt=1.5 ps and $\tau_B$ = 46.7 ps).

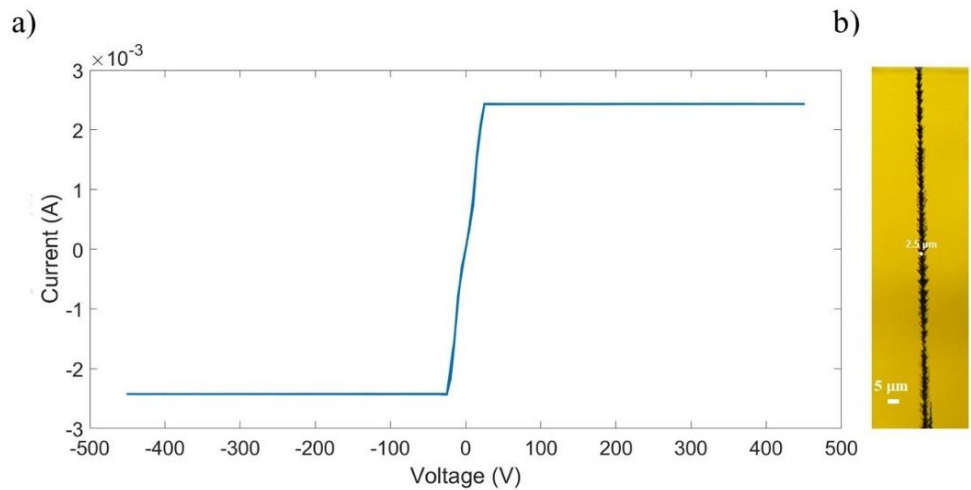

Figure 5: a) IV graph of a micro electrode fabricated in diamond by Bessel beam machining with laser bursts featured by n=32 sub-pulses, $\tau_B$ = 46.7 ps, and Δt =1.5 ps (mode 1) at 20 Hz and with 10 µJ burst energy.

Note that the horizontal parts of the curve are due to the reaching of the compliance current of the electrometer set at 2.5 mA. The measured resistance is 10.4 kΩ. b) Zoomed optical microscope image of the central portion of the corresponding graphitic wire.

Note that while in this particular burst mode writing regime the measured resistivity values turn out to be low (<0.03 Ω cm) throughout the different burst energy values considered, the effect of annealing is in fact negligible (as shown in table 2). This is in accordance with what previously observed for the graphitic electrodes fabricated in the single pulse laser writing regime, and featured already by low resistivity before annealing [26].

## 3.4 Effect of thermal annealing

The role of thermal annealing requires a specific attention here. One of the main targets of annealing of the diamond sample is to reduce the potential barrier, if present, of the IV curve of the electrodes fabricated. In the work presented in [26], we have shown that the only entity where annealing seems to have an effect is the resistivity of the laser fabricated graphitic wires rather than the barrier height itself. In here, we report for the first time, a reduction of the barrier potential that has been observed in the IV curve of some Bessel written electrodes with laser bursts, after annealing of the diamond sample. It is worth noting that in fact in the burst mode writing regime very few electrodes have shown the presence of a barrier potential irrespective of the fabrication parameters. An example is given by a graphitic wire fabricated at 20 Hz with laser bursts of duration $\tau_B$ = 46.7 ps containing n = 2 sub-pulses (mode 2) using 8 µJ per burst. In Figure 6 we report the IV curves measured for that micro electrode and showing the disappearance of the 10 V barrier potential after the thermal annealing. Moreover, after the annealing process we can observe a decrease in the resistivity from 0.088 Ω cm to 0.076 Ω cm.

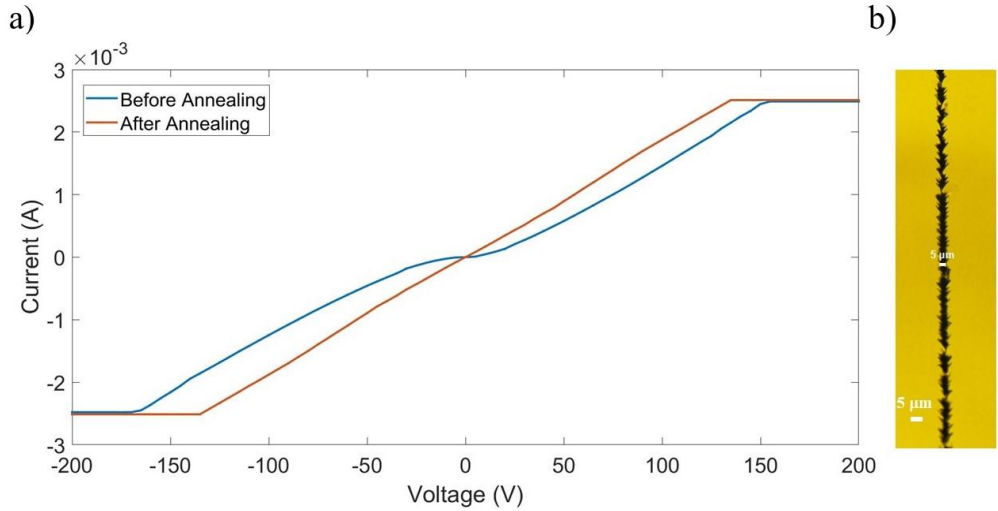

Figure 6: a) IV graphs of a micro electrode fabricated at 20 Hz in diamond by Bessel beam machining with laser bursts featured by n= 2 and $\tau_B$ = 46.7 ps (mode 2) with 8 µJ per burst, and measured before (blue curve) annealing and after (red curve) annealing respectively. b) Zoomed optical microscope image of the central portion of the corresponding graphitic wire (the morphology is the same before or after annealing).

The difference before and after thermal annealing in the electrical response of the electrodes with respect to the presence or absence of a potential barrier and in their conductivity may be explained on the basis of how the transformation of diamond into $sp^2$ carbon occurs in each case. As commented in [26], the presence of a potential barrier in the IV curve denotes the presence of micro/nano spacing between different graphitic globules/sheets formed during the graphitization process of the diamond material when the Bessel beam interacts with the material and, as reported in our previous work, the higher the fabrication energy, the smaller might be these gaps and thus, the lower the barrier height during the electrical current flow. Moreover, this trend is common irrespective of the laser writing regime (single pulses or bursts). When it comes to annealing, the electrical response of the electrodes is different since those fabricated by using laser bursts show a difference in both barrier height and conductivity while those fabricated in the standard single pulse mode machining regime show a difference only in conductivity. One possible explanation could be that in the case of graphitic wires machined with bursts, even though there are micro or nano gaps between the graphitic sheets, the material in those spaces might not be just pure $sp^3$ diamond but it could have been already modified by the heat induced by the laser bursts in such a way that it is already at the graphitization threshold. However, those gaps are not yet conductive. Therefore, during the annealing, they might undergo an efficient transformation into $sp^2$ hybridisation as in case of electrodes fabricated with ion implantation technique where diamond is converted into a stable graphitic phase upon thermal annealing [48]. This results in the partial or complete removal of the micro gaps within the graphitic globules depending on the energy of the laser modification. Moreover, the same effect can be seen in the regions which are already transformed but in a different manner. The regions of the crystalline material which are already modified, even if only partially, can profit from the annealing to complete the transformation and increase their graphite content, with a consequent reduction of the resistivity of the micro electrode. Therefore, one could safely assume that there are two kinds of transformations that occur in case of graphitic wires fabricated with laser bursts while annealing the diamond sample: a) conversion of the region/gap which is at graphitisation threshold into a conductive region leading to a reduction in the barrier height and b) conversion of the partially transformed regions that are already above the graphitisation threshold into a fully transformed $sp^2$ phase leading to an increase in conductivity. The former mechanism may not be a complete transformation and instead it could be only partial. However, it still can become conductive leading to reduction in the barrier height. Note that for the graphitic wires fabricated using single pulse writing regime, only the second type of material transformation occurs, leading in that case to an increase in conductivity but no reduction of the potential barrier (see [26]).

## *3.5 Burst vs single-mode laser writing*

The difference in the response to thermal annealing of electrodes fabricated using laser bursts or in the standard single pulse mode-regime, establishes the fact that the current-voltage characteristics can change with respect to the laser fabrication mode. Therefore, a comparison between the conductivities of electrodes fabricated respectively with bursts and single pulses featured by the same total duration and energy has been established herewith. We recall that in the burst the energy is equally split among the sub-pulses. Figure 7 schematizes the two cases: a burst containing fs sub-pulses and having a given total duration 10.7 ps, and a single laser pulse with approximately the same temporal duration (10 ps).

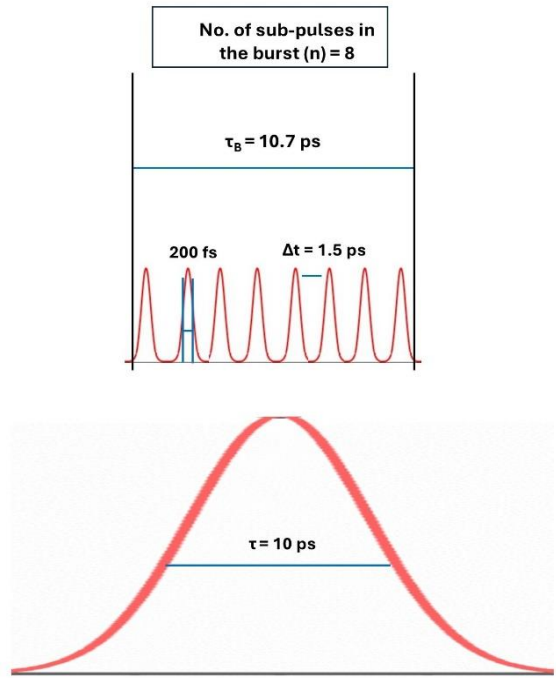

Figure 7: (Top) schematic of a burst in mode 3 (a sequence of very compact 200 fs n=8 sub-pulses within the ≈10 ps duration with distributed energy among them) and (bottom) the standard Gaussian pulse mode (with 10 ps duration).

For a comparison of the radiation-matter interaction occurring in diamond in the two equivalent laser writing configurations above mentioned (equivalent for what concerns the duration of the light package arriving on the sample, either bursts or single pulses), we have analysed the difference in the IV graphs characterizing the graphitic micro wires obtained. In Figure 8 we report the IV curves respectively measured for an electrode fabricated at 20 Hz with 10.7 ps long bursts containing 8 sub-pulses of duration 200 fs (mode 3) and for an electrode fabricated with 10 ps single pulses in the case of laser machining with burst/single pulse energy of 10 µJ.

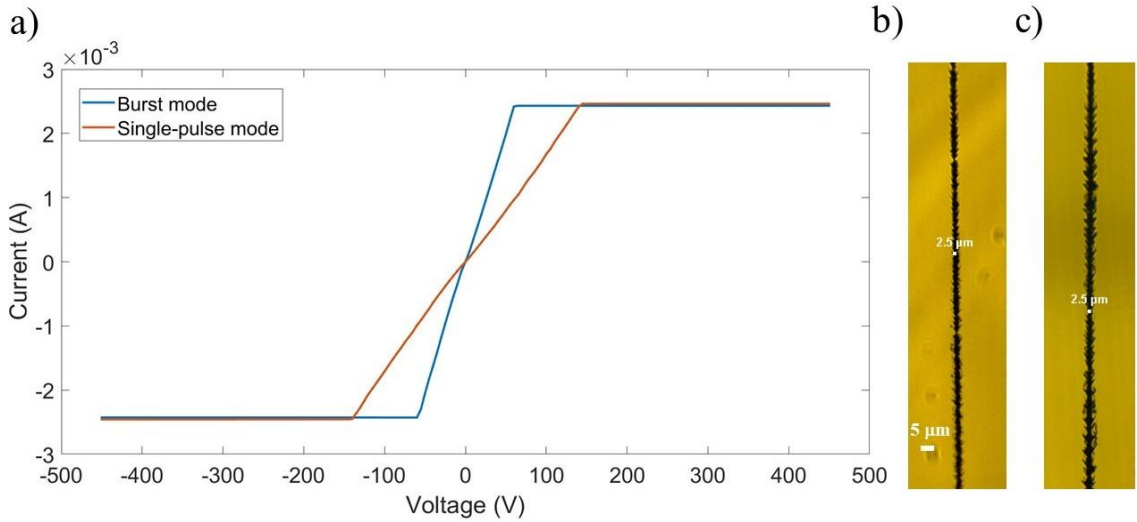

Figure 8: a) IV graphs of micro electrodes fabricated in two different laser writing regimes, but with equivalent single pulse/burst duration; i.e. with laser bursts with n = 8 and $\tau_B$ = 10.7 ps (mode 3) (blue curves) and with a train of 10 ps single pulses (red curves) having same burst/pulse energy equal to 10 µJ. The measured resistance values are 25 kΩ and 58 kΩ respectively. The zoomed optical microscope images refer to the central portions of the graphitic wires obtained respectively with bursts (b) and with a train of single pulses (c). The scale bar is the same for both images.

From the slope of the ohmic portions of the graphs one can appreciate the difference between the resistivity of the micro electrodes fabricated in the standard single pulse mode regime, and those fabricated with laser bursts. While in the latter case, the resistivity turns out to be very low (≈ 0.02 Ω cm), in the single pulse writing mode the resistivity value measured is given by 0.08 Ω cm. Such behaviour may be ascribed to the peculiar cumulative damage process occurring during the burst mode processing of materials [48]. In fact, the damage-induced graphite globules accumulate sub-pulse after sub-pulse within the bursts, possibly resulting in a better energy absorption and leading to much more efficient graphitization process compared to single pulse irradiation.

We report for completeness the difference in the measured micro-Raman spectra related for instance to the graphitic wires fabricated in the two laser writing configurations (single pulses and bursts respectively) with 10 µJ pulse/burst energy, in order to analyse a possible difference in the sp$^2$ carbon formation. Figure 9 shows the spectra obtained (black curve) and the "deconvolved" peaks considering the contributions of diamond and of amorphous carbon.

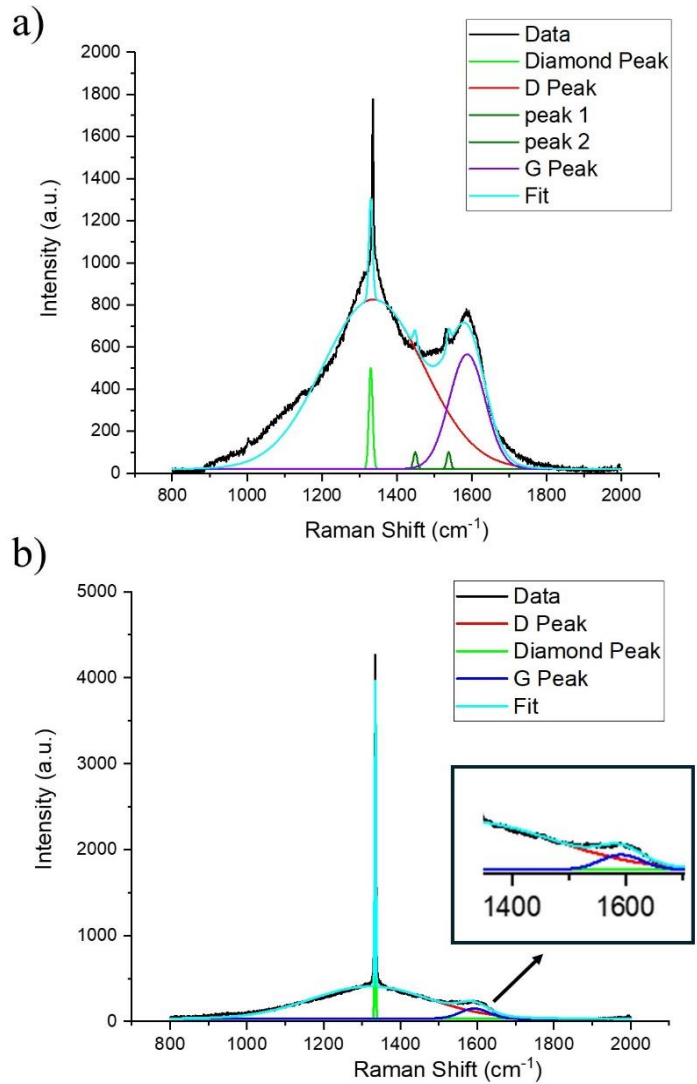

Figure 9: Micro-Raman spectra of two graphitic wires fabricated at 20 Hz with Bessel beams using a) bursts with n = 8 and $\tau_B$ = 10.7 ps (mode 3) and b) single pulses with approximately the same duration (10 ps). The pulse/burst energy was 10 μJ.

The obtained results related to the two writing configurations, i.e burst mode laser writing (Fig. 9a) and single pulse laser writing (Fig. 9b), clearly suggest that there is difference in the amount of transformation of $sp^3$ carbon into $sp^2$ carbon occurring in the two different cases. In both cases the presence of the additional D and G bands together with the diamond peak indicates that the diamond crystalline structure is modified. But the sharper G band and the presence of additional two bands (shown in dark green in Fig. 8a) at 1447 $cm^{-1}$ and 1528 $cm^{-1}$ measured for the micro wire generated in the burst fabrication mode suggest that there is more conductive material in this electrode compared to the one fabricated in the single pulse mode. Moreover, the ratio of the intensity of the G band to the diamond peak of the electrode fabricated with bursts is around 10 times greater than that of the electrode fabricated with single pulses pointing out the better graphitization process in the first case.

# 4. Graphitic electrode fabrication with picosecond sub-pulses

In this last section, we summarize for completeness the results relative to the graphitic micro wires fabricated by Bessel beams in diamond when using laser bursts featured by picosecond sub-pulses, even if only two sub-pulses were contained within the bursts. The bursts configurations considered are those featured respectively by n=2 with $\tau_B$ = 56.5 ps, $\Delta t$= 46.5 ps (mode 5), and by n=2 with $\tau_B$ = 46 ps, $\Delta t$=36 ps (mode 6). In both configurations the sub-pulses within the bursts have 10 ps duration. Our experiments have been performed by using three burst energies, and at two laser repetition rates, namely 20 Hz and 200 kHz.

The most interesting result we have observed (and different result with respect to the fs sub-pulses case) is that the micro electrodes fabricated in the picosecond sub-pulse regime of the laser bursts tend to be featured by an almost constant low resistivity, over the whole range of laser writing parameters and configurations used. This is seen by the fact that the lowest resistivity we have found is 0.014 $\Omega$ cm and the highest resistivity is a mere resistivity value of 0.026 $\Omega$ cm (these values being the same for electrodes fabricated at 20 Hz or 200 kHz). Moreover, none of the electrodes fabricated with bursts featured by picosecond sub-pulses showed any evident potential barrier in the IV electrical curves, irrespective of the laser writing parameters (energy and repetition rate). In Figure 10 we report as an example the IV graphs relative to two micro electrodes fabricated at 20 Hz repetition rate in the two above-mentioned burst mode writing configurations (mode 5 and mode 6 respectively). The micro wires whose IV curves are reported in the upper row have been fabricated with 8 µJ burst energy, while those related to the results of the lower row have been fabricated with 6 µJ burst energy. All the resulting graphs are very similar.

In general, from the Bessel beam machining experiments where different combinations of burst energies and repetition rates have been used, all the graphitic micro wires fabricated in the (100) cut diamond samples with laser bursts containing two sub-pulses of 10 ps duration, have shown to be conductive with resistivity lying in the range 0.014 - 0.02 $\Omega$ cm.

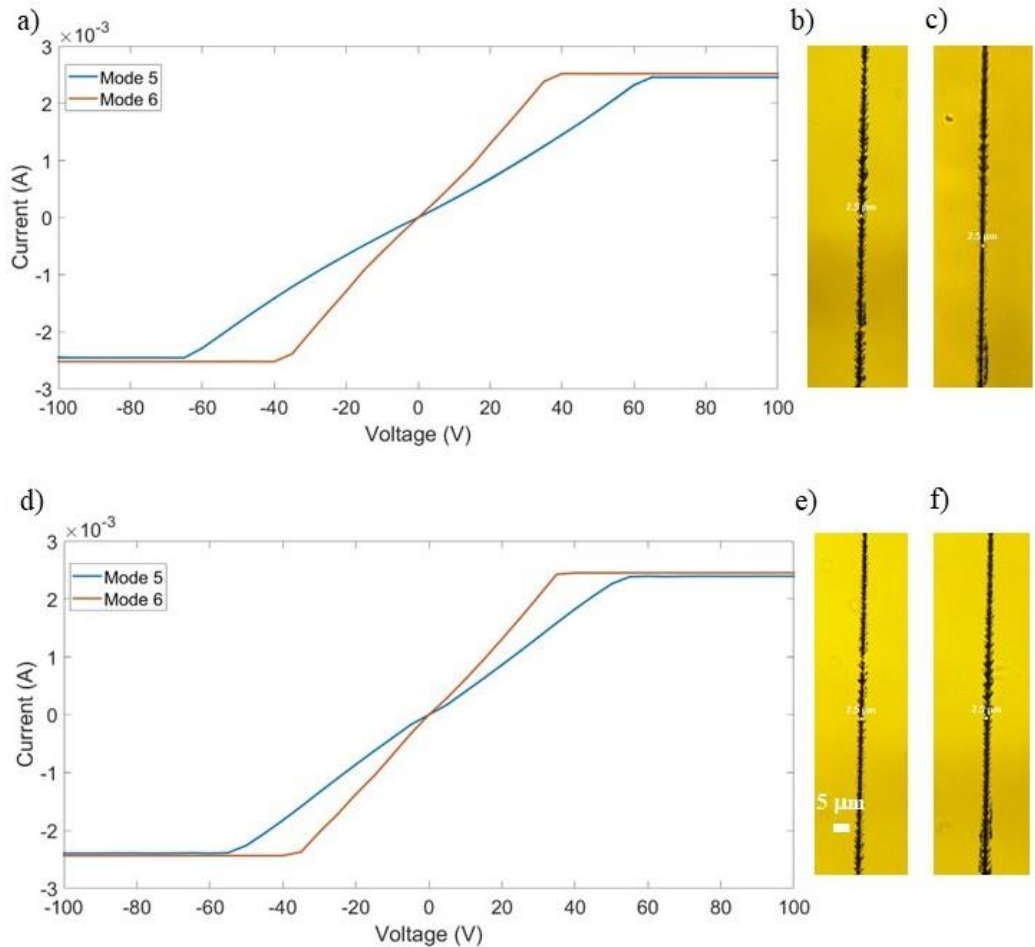

Figure 10: IV characteristics of electrodes fabricated by Bessel beams with laser bursts featured by 10 ps sub-pulses using a) 8 µJ and d) 6 µ energy per burst at 20 Hz repetition rate. The burst configurations used are those denoted respectively by mode 5 where the bursts are featured by n = 2 with $\tau_B$ = 56.5 ps (the resistance is on the order of ~15 kΩ) and by mode 6 with n = 2 with $\tau_B$ = 46 ps. The zoomed optical microscope images refer to the central portion of the graphitic wires obtained with bursts of mode 5 (b and e) featured by resistance ~25 kΩ and with bursts of mode 6 (c and f) respectively, featured by resistance ~15 kΩ. The scale bar is the same for all images.

Before concluding we shall make a final comment on the fact that in this work we deliberately chose to compare different writing modes and to present the results for a fixed pulse/burst energy, chosen in a range where the fabricated micro electrodes showed the best conductivities. By reviewing all the experimental conditions explored we could nevertheless observe in two experimental cases where the graphitic wires fabricated at 20 Hz with single pulses and with laser bursts had approximately the same peak power, that the burst fabrication mode leads in any case to lower resistivity values with respect to the single pulse writing mode. In particular, by comparing the results obtained with 46.7 ps long, 8 μJ, bursts of mode 2 (n=2, 200 fs sub pulses) and single 200 fs pulses of 6 μJ (peak powers on the order of ~ 2-3 $10^7$ W) we found a resistivity of 0.076 Ω cm in the first case and 0.43 Ω cm in the second. Similarly, by comparing the results obtained with 46 ps long, 10 μJ bursts of mode 6 (n=2, 10 ps sub pulses) and single 10 ps pulses of 6 μJ (peak powers on the order of 5. $10^5$ W) the resistivity of the wire fabricated in the burst mode turned out to be 0.01 Ω cm while that of the wire fabricated in single pulse mode was about 0.02 Ω cm.

# 5. Conclusions

In this work we have reported on the microfabrication of in-bulk conductive graphitic wires perpendicular to the sample surface of a 500 μm thick monocrystalline HPHT type IIa diamond using pulsed Bessel beams in the burst laser mode regime. The duration of the laser pulses and thus also of the sub-pulses within the bursts was set to 200 fs and to 10 ps for comparison. The obtained micro electrodes have been electrically characterized by current-voltage tests, and micro-Raman spectroscopy has been used to analyse the structural features of the micro wires.

Different available burst configurations (featured by different burst duration, number of sub-pulses within the burst, and intra-burst time delay) have been used for the fabrication of the graphitic wires in diamond. In the case of laser bursts with 200 fs sub-pulses, the configuration with the highest number of sub-pulses (i.e. mode 1 with 32 sub-pulses), lowest time delay (1.5 ps, i.e. sub THz regime) and longest total burst duration (46.7 ps) led to the best conductivity (lowest evaluated resistivity) compared to other configurations, both at 20 Hz and 200 kHz repetition rates. We believe that with a high number of closely packed sub-pulses within the bursts even if the total energy delivered to the material does not change but rather it is split equally between the sub-pulses in the burst, the heating process in the material during the radiation-matter interaction is more efficient (thanks to the reduced leakage of heat) compared to that occurring with a standard train of single pulses of the same duration as the total burst duration (or to that of a smaller number of pulses within the burst). As a consequence, we expect to have better transformation of diamond into $sp^2$ carbon and thus, better micro electrodes with lower resistivity value. In addition, a burst duration of tens of picosecond could also contribute to this efficient transformation. In particular in that case, the progressive graphitic damage accumulation that occurs during the repetitive irradiation with 200 fs sub-pulses within the burst may be responsible for the more efficient graphitic formation with respect to the case of the arrival of single pulses of 10 ps duration. Moreover, the electrodes fabricated with laser bursts (in the fs sub-pulse regime) delivered at 20 Hz showed better conductivity than those fabricated with a laser repetition rate of 200 kHz. The same has been verified using micro-Raman measurements which revealed better transformation of $sp^3$ hybridized carbon into $sp^2$ hybridization in the first case.

While no trace of evident potential barrier was observed in the electrical characterization of the micro electrodes fabricated with bursts featured by 10 ps sub-pulses, we noticed a reduction in the potential barrier after thermal annealing of the sample, in the IV curves of the electrodes fabricated using sub-pulses of 200 fs duration within the burst. To the best of our knowledge, this is the first time a study reports a variation in the potential barrier with respect to thermal annealing for laser written electrodes. Moreover, the resistivity of the graphitic micro electrodes fabricated by using Bessel beams in the burst mode regime also gets reduced after annealing. This may happen because of two transformation mechanisms of the diamond crystalline structure during the laser-matter interaction, namely, first the conversion of the micro/nanogaps (which are at graphitisation threshold) into a conductive region leading to a reduction in the barrier height; and, second, the conversion of the partially transformed region (already above graphitisation threshold) into a fully transformed $sp^2$ phase leading to an increase in conductivity.

Finally, in experiments aiming at comparing the radiation-matter interaction occurring in diamond in the standard writing mode and in the burst writing mode, featured by similar pulse/burst durations ($\approx$ 10 ps), we clearly observed the advantage of using laser bursts for the fabrication of graphitic micro wires with very low resistivity (and lower than that measured in electrodes fabricated with standard train of single pulses). This is also in accordance with the results of the micro-Raman spectroscopy revealing the presence of a sharp G band and two additional spectral peaks estimated to favour the electrical conductivity of the micro electrodes fabricated with bursts.

To conclude, through a careful optimization of the laser writing parameters (pulse energy and repetition rate), and of the laser burst configurations, we have been able to obtain micro electrodes featured by resistivities as low as of 0.012 $\Omega$ cm when using 200 fs sub-pulses and 0.014 $\Omega$ cm when using 10 ps sub-pulses within the burst. In other words, the micromachining of diamond in burst mode writing regime allows to reach low resistivity values irrespective of the laser source pulse duration, thus challenging the general notion that larger pulse durations (for instance the use of picosecond pulses instead of femtoseconds) gives better conductivity of the graphitic wires. At the same time, our experiments showed that, while in the femtosecond laser writing regime the results are more variable, the micromachining process in the picosecond regime (i.e. using bursts with two sub-pulses of 10 ps duration) allows to generate micro electrodes with resistivity levels lying in the range of 0.014 – 0.026 Ω cm irrespective of the pulse energy and repetition rate used.

The reported resistivity values in this work are found to be, to the best of our knowledge, among the lowest if not the lowest for what concerns laser written graphitic electrodes in the literature, and these are quite impressive if we compare them with the typical resistivity values measured for standard polycrystalline graphite ($\rho \approx 3.5 \times 10^{-3}$ Ω cm) [48]. Moreover, the absence of barrier potential in the electrical features of the micro electrodes obtained with Bessel beam burst mode machining, makes such electrodes of good quality. We thus believe that the graphitic wires created with our technique can find their role in diamond-based applications such as photonic circuits for NV sensing, microfluidic chips, and possibly also in radiation detectors for particle physics and medical dosimetry where in-bulk, highly conductive graphitic electrodes are required.

## Authorship contribution statement

O. J. conceived, supervised the micromachining experiments, discussed the results and revised the first manuscript draft. A. K., C. G. and F. M. realized the micromachining of the diamond samples, under the supervision of A. A.     A. K. analysed the results, and wrote the first manuscript draft. Raman measurements were performed and analysed by A. C and A. K. The electrical measurements were performed by A. K under the supervision of F. P. All authors revised the manuscript.

## Declaration of competing interests

The authors declare no conflict of interest.

## Acknowledgements

The authors would like to thank Vanna Pugliese for helping with the annealing processes, Pietro Aprà and Adam Britel for helping with the electrical characterizations, and Paolo Olivero, from the University of Torino, for useful discussions.

This research has received funding from the European Union's H2020 Marie Curie ITN project LasIonDef (GA n.956387) and QuantDia (FISR2019-05178) funded by Ministero dell'Istruzione, dell'Università e della Ricerca.

# References

1. C. Bloomer, M. E. Newton, G. Rehm, P. S. Salter. **A single-crystal diamond X-ray pixel detector with embedded graphitic electrodes**. Journal of Synchrotron Radiation 27, 3 (2020), 599-607. https://doi.org/10.1107/S160057752000140X

2. Bharadwaj, O. Jedrkiewicz, J. P. Hadden, B. Sotillo, M. R. Vázquez, P. Dentella, T. T. Fernandez, A. Chiappini, A. N. Giakoumaki, T. L. Phu, M. Bollani, M. Ferrari, R. Ramponi, P. E. Barclay, S. M. Eaton, 2019. **Femtosecond laser written photonic and microfluidic circuits in diamond.** Journal of Physics: Photonics 1, 022001. https://doi.org/10.1088/2515-7647/ab0c4e

3. B. Sotillo, V. Bharadwaj, J. P. Hadden, M. Sakakura, A. Chiappini, T. T. Fernandez, S. Longhi, O. Jedrkiewicz, Y. Shimotsuma, L. Criante, R. Osellame, G. Galzerano, M. Ferrari, K. Miura, R. Ramponi, P. E. Barclay, S. M. Michael Eaton, 2016. **Diamond photonics platform enabled by femtosecond laser writing**. Scientific Reports 6, 35566. https://doi.org/10.1038/srep35566

4. Handbook of Industrial Diamonds and Diamond Films, Marcel Dekker, New York, 1998.

5. J. Tapper. **Diamond detectors in particle physics**. Reports on Progress in Physics 63, 8 (2000), 1273–1316. https://doi.org/10.1088/0034-4885/63/8/203

6. V. Carabelli, A. Marcantoni, F. Picollo, A. Battiato, E. Bernardi, A.Pasquarelli, P.Olivero, and E. Carbone, **Planar Diamond-Based Multiarrays to Monitor Neurotransmitter Release and Action Potential Firing: New Perspectives in Cellular Neuroscience**, ACS Chemical Neuroscience 8, 2 (2017), 252–264. https://doi.org10.1021/acschemneuro.6b00328

7. . V. R. Howes. **The graphitization of diamond**, Proceedings of the Physical Society 80 (1962) 648–662. https://doi.org/10.1088/0370-1328/80/3/310

8. A. Battiato, M. Lorusso, E. Bernardi, F. Picollo, F. Bosia, D. Ugues, A. Zelferino, A. Damin, J. Baima, N.M. Pugno, E.P. Ambrosio, P. Olivero, **Softening the ultra-stiff: Controlled variation of Young's modulus in single-crystal diamond by ion implantation**, Acta Materiala, 116 (2016) 95-103. https://doi.org/10.1016/j.actamat.2016.06.019

9. F Picollo, D Gatto Monticone, P Olivero, B A Fairchild, S Rubanov, S Prawer and E Vittone, **Fabrication and electrical characterization of three-dimensional graphitic microchannels in single crystal diamond**, New Journal of Physics 14 053011 (2012). https://doi.org/10.1088/1367-2630/14/5/053011

10. K. K. Ashikkalieva, **Laser-Induced Graphitization of Diamond Bulk: The State of the Art (A Review)**, Physics of Wave Phenomena 30 (2022), 1-16. https://doi.org/10.3103/S1541308X22010034

11. T. V. Kononenko, E. V. Zavedeev, V. V. Kononenko, K. K. Ashikkalieva, V. I. Konov. **Graphitization wave in diamond bulk induced by ultrashort laser pulses**, Applied Physics A 119 (2015), 405-414. https://doi.org/10.1007/s00339-015-9109-0

12. Y. Shimotsuma, M. Sakakura, S. Kanehira, J. Qiu, P. G. Kazansky, K. Miura, K. Fujita, K. Hirao, **Three-dimensional nanostructuring of transparent materials by the femtosecond laser irradiation**, J. Laser Micro Nanoeng. 1 (2006) 181-184. https://doi.org/10.2961/jlmn.2006.03.0006

13. M. Shimizu, Y. Shimotsuma, M. Sakakura, T. Yuasa, H. Homma, Y. Minowa, K. Tanaka, K. Miura, K. Hirao, **Periodic metallo-dielectric structure in diamond**, Opt. Express 17 (2009) 46-54. https://doi.org/10.1364/OE.17.000046

14. T.V. Kononenko, M.S. Komlenok, V.P. Pashinin, S.M. Pimenov, V.I. Konov, M. Neff, V. Romano, W. Lüthy, **Femtosecond laser microstructuring in the bulk of diamond**, Diamond Relat. Mater. 18 (2009) 196-199. https://doi.org/10.1016/j.diamond.2008.07.014

15. R. R. Gattass, E. Mazur. **Femtosecond laser micromachining in transparent materials**, Nature Photonics 2 (2008), 219-225. https://doi.org/10.1038/nphoton.2008.47

16. B. Sun, P. S. Salter, and M. Booth, **High conductivity micro-wires in diamond following arbitrary paths**, Appl. Phys. Lett. 105 (2014), 231105. https://doi.org/10.1063/1.4902998

17. J. Durnin, J. J. Miceli, Jr., and J. H. Eberly. **Diffraction free beams**. Physical Review Letters 58, 1499 (1987). https://doi.org/10.1103/PhysRevLett.58.1499

18. J. Durnin, J. J. Miceli, J. H. Eberly, **Comparison of Bessel and Gaussian beams**, Optics Letters 13 (1988), 79-80. https://doi.org/10.1364/ol.13.000079

19. M. Duocastella, C.B. Arnold, **Bessel and annular beams for materials processing**, Laser and Photonics Reviews 6 (2012), 607-621. https://doi.org/10.1002/lpor.201100031

20. V. Garzillo, V. Jukna, A. Couairon, R. Grigutis, P. di Trapani, O. Jedrkiewicz, **Optimization of laser energy deposition for single-shot high aspect-ratio microstructuring of thick BK7 glass**. Journal of Applied Physics. 120. 013102 (2016). https://doi.org/10.1063/1.4954890

21. M. K. Bhuyan, O. Jedrkiewicz, V. Sabonis, M. Mikutis, S. Recchia, A Aprea, M. Bollani, P. Di Trapani, **High-speed laser-assisted cutting of strong transparent materials using picosecond Bessel beams**, Applied Physics A 120 (2015), 443-446. http://dx.doi.org/10.1007/s00339-015-9289-7

22. R. Stoian, M. K. Bhuyan, G. Zhang, G. Cheng, R. Meyer, F. Courvoisier, **Ultrafast Bessel beams: advanced tools for laser materials processing**, Advanced Optical Technologies 7 (2018), 165–174. https://doi.org/10.1515/aot-2018-0009

23. A. Kuriakose, M. Bollani, P. Di Trapani, O. Jedrkiewicz, 2022. **Study of Through-Hole Micro-Drilling in Sapphire by Means of Pulsed Bessel Beams**. Micromachines 13(4), 624. https://doi.org/10.3390/mi13040624

24. R. Meyer, M. Jacquot, R. Giust, J. Safioui, L. Rapp, L. Furfaro, P.-A. Lacourt, J. M. Dudley, F. Courvoisier, **Single-shot ultrafast laser processing of high-aspect-ratio nanochannels using elliptical Bessel beams**, Optics Letters 42 (2017), 4307-4310. https://doi.org/10.1364/OL.42.004307

25. A. Kuriakose, A. Chiappini, B. Sotillo, A. Britel, P. Aprà, F. Picollo, O. Jedrkiewicz. **Fabrication of conductive micro electrodes in diamond bulk using pulsed Bessel beams.** Diamond and Related Materials, Volume 136 (2023). 110034. https://doi.org/10.1016/j.diamond.2023.110034

26. A. Kuriakose, A. Chiappini, P. Aprà, O. edrkiewicz. **Effect of crystallographic orientation on the potential barrier and conductivity of Bessel written graphitic electrodes in** **diamond**. Diamond & Related Materials, Volume 142 (2024) 110760. https://doi.org/10.1016/j.diamond.2023.110760

27. V.V. Belloni, V. Sabonis, P. Di Trapani, O. Jedrkiewicz, 2020. **Burst mode versus single-pulse machining for Bessel beam micro-drilling of thin glass: study and comparison**. SN Applied Sciences 2, 1589. https://doi.org/10.1007/s42452-020-03327-4

28. P. R. Herman, M. Lapczyna, D. Breitling, H. Schittenhelm, H.W. Tan, Y. Kerachian, R. S. Marjoribanks, **Laser micromachining of transparent glasses and aluminum with ps-pulse bursts at 1054 nm,** Conference on Lasers and Electro-Optics (2000). https://opg.optica.org/abstract.cfm?URI=CLEO-2000-CFD3

29. A. T. Ellis, M. E. Fourney, **Application of a ruby laser to high-speed photography**, Proceedings of IEEE 51 (1963), 942-943. https://doi.org/10.1109/PROC.1963.2340

30. S. Rezaei, J. Li, P. R. Herman, **Burst train generator of high energy femtosecond laser pulses for driving heat accumulation effect during micromachining**, Optics Letters 40 (2015), 2064-2067. https://doi.org/10.1364/OL.40.002064

31. M. Lapczyna, K. P. Chen, P. R. Herman, H. W. Tan, and R. S. Marjoribanks, **Ultra high repetition rate (133 MHz) laser ablation of aluminum with 1.2-ps pulses**, Applied Physics A 69 (1999), 883-886. http://dx.doi.org/10.1007/s003390051552

32. S. Döring, T. Ullsperger, F. Heisler, S. Richter, A. Tünnermann, and S. Nolte, **Hole Formation Process in Ultrashort Pulse Laser Percussion Drilling**, Physics Procedia 41 (2013), 431-440. https://doi.org/10.1016/j.phpro.2013.03.099

33. C. Gaudiuso, F. Fanelli, F. P. Mezzapesa, A Volpe, A. Ancona, 2023. **Tailoring the wettability of surface-textures copper using sub-THz bursts of femtosecond laser pulses**, Applied Surface Science 638, 158032. https://doi.org/10.1016/j.apsusc.2023.158032

34. D. J. Forster, B. Jaggi, A. Michalowski, B. Neuenschwander, 2021. **Review on Experimental and Theoretical Investigations of Ultra-Short Pulsed Laser Ablation of Metals with Burst Pulses**, Materials 14, 3331. https://doi.org/10.3390/ma14123331

35. T. Kramer, Y. Zhang, S. Remund, B. Jaeggi, A. Michalowski, L. Grad, B. Neuenschwander, 2017. **Increasing the Specific Removal Rate for Ultra Short Pulsed Laser-Micromachining by Using Pulse Bursts**, Journal of Laser Micro/Nanoengineering 12, 2. https://doi.org/10.2961/jlmn.2017.02.0011

36. C. Gaudiuso, B. Stampone, G. Trotta, A. Volpe, A. Ancona, 2023. **Investigation of the micro-milling process of steel with THz bursts of ultrashort laser pulses**, Optics & Laser Technology 162, 109301. https://doi.org/10.1016/j.optlastec.2023.109301

37. Kuriakose, M. Bollani, P. Di Trapani, O. Jedrkiewicz, **Study of through-hole micro-drilling in sapphire by means of pulsed Bessel beams**, Micromachines 13 (4) (2022) 624, https://doi.org/10.3390/mi13040624

38. Dromey B, Zepf M, Landreman M, O'keeffe K, Robinson T, Hooker SM. **Generation of a train of ultrashort pulses from a compact birefringent crystal array**, Appl Opt (2007) 46:5142–6. https://doi.org/10.1364/AO.46.005142

39. S. Zhou, D. Ouzounov, H. Li, I. Bazarov, B. Dunham, C. Sinclair, and F. W. Wise, **Efficient temporal shaping of ultrashort pulses with birefringent crystals**, Appl. Opt. **46**(35), 8488–8492 (2007); https://doi.org/10.1364/AO.46.008488

40. K. Mishchik, R. Beuton, O. Dematteo Caulier, S. Skupin, B. Chimier, G. Duchateau, B. Chassagne, R. Kling, C. Hönninger, E. Mottay, and J. Lopez, **Improved laser glass cutting by spatio-temporal control of energy deposition using bursts of femtosecond pulses**, Optics Express 25 (2017), 33271-33282. https://doi.org/10.1364/OE.25.033271

41. B. Neuenschwander, B. Jaeggi, E. V. Zavedeev, N. R. Arutyunyan, S. M. Pimenov, (2019). **Heat accumulation effects in laser processing of diamond-like nanocomposite films with bursts of femtosecond pulses.** Journal of Applied Physics, 126(11), 115301. https://doi.org/10.1063/1.5121424

42. Anna Dychalska, P. Popielarski, Wojciech Frankow, ´K. Fabisiak, K. Paprocki, Mirosław Szybowicz, **Study of CVD diamond layers with amorphous carbon admixture by Raman scattering spectroscopy**, Mater. Sci.-Pol. 33 (2015) 799–805, https://doi.org/10.1515/msp-2015-0067.

43. A. Karczemska, M. Szurgot, M. Kozanecki, M.I. Szynkowska, V. Ralchenko, V.V. Danilenko, P. Louda, S. Mitura, **Extraterrestrial, terrestrial and laboratory diamonds — Differences and similarities**, Diamond and Related Materials, Volume 17, 7–10(2008), Pages 1179-1185, https://doi.org/10.1016/j.diamond.2008.02.021

44. A. Kromka, J. Breza, M. Kadlecikova, J. Janik, F. Balon, **Identification of Carbon Phases and Analysis of Diamond/Substrate Interfaces by Raman Spectroscopy**, Carbon 43, 2(2005), 425–429. https://doi.org/10.1016/j.carbon.2004.10.004

45. J. Birrell, J.E. Gerbi, O. Auciello, J.M. Gibson, J. Johnson, J.A. Carlisle, **Interpretation of the Raman spectra of ultrananocrystalline diamond**, Diamond and Related Materials, Volume 14, 1(2005), 86-92. https://doi.org/10.1016/j.diamond.2004.07.012

46. A.C. Ferrari, J. Robertson, 2001. **Origin of the 1150 cm$^{-1}$ Raman Mode in Nanocrystaalline Diamond**, Physical Review B 63 121405. https://doi.org/10.1103/PhysRevB.63.121405

47. F. Banhart, **Elemental carbon in the sp1 hybridization**. ChemTexts 6, 3(2020). https://doi.org/10.1007/s40828-019-0098-z

48. C. Gaudiuso, G. Giannuzzi, A. Volpe, P. M. Lugarà, I. Choquet, and A. Ancona, **Incubation during laser ablation with bursts of femtosecond pulses with picosecond delays**, Optics Express 26(4), 2018. https://doi.org/10.1364/OE.26.003801

49. F. Picollo, D. Gatto Monticone, P. Olivero, B. A. Fairchild, S. Rubanov, S. Prawer, E. Vittone, 2012. **Fabrication and electrical characterization of three-dimensional graphitic

**microchannels in single crystal diamond**, New Journal of Physics 14, 053011. https://doi.org/10.1088/1367-2630/14/5/053011